\documentclass[showpacs,preprintnumbers,amsmath,aps,amssymb]{revtex4}
\usepackage{graphicx}
\usepackage{dcolumn}
\usepackage{bm}
\begin{document}

\baselineskip 17pt

\title{Scalar field in the  Bianchi I: Non commutative classical and  Quantum Cosmology }
\author{J. Socorro$^{1,2}$}
\email{socorro@fisica.ugto.mx}
\author{Luis O. Pimentel$^3$}
\email{lopr@xanum.uam.mx}
\author{C. Ortiz$^{1,4}$}
\email{ortizgca@fisica.uaz.edu.mx}
\author{M. Aguero$^2$}
\email{makxim@gmail.com}

\affiliation{$^1$Departamento de  F\'{\i}sica, DCI-Campus Le\'on, Universidad de Guanajuato,\\
A.P. E-143, C.P. 37150, Le\'on, Guanajuato, M\'exico\\
$^2$Facultad de Ciencias de la Universidad Aut\'onoma del Estado de
M\'exico, \\
Instituto Literario  No. 100, Toluca, C.P. 50000, Edo de Mex, M\'exico \\
$^3$Departamento de F\'{\i}sica, Universidad Aut\'onoma Metropolitana,
Apartado Postal 55-534, 09340, M\'exico, D.F.\\
$^4$Unidad Acad\'emica de F\'{\i}sica, Universidad Aut\'onoma de Zacatecas\\
Cda. Solidaridad SN, CP. 98060, Zacatecas, Zac., M\'exico}%

\date{\today}

\begin{abstract}
Using the ADM formalism in the minisuperspace, we obtain the commutative and
noncommutative exact classical solutions and exact wave function
to the Wheeler-DeWitt equation with an arbitrary factor ordering, for the
anisotropic Bianchi type I  cosmological model, coupled to a scalar field,
cosmological term and barotropic perfect fluid.
We introduce noncommutative scale factors,  considering that  all
minisuperspace  variables $\rm q^i$ do not commute,
so the symplectic structure was modified. In the classical regime, 
it is shown that the anisotropic parameter $\rm \beta_{\pm nc}$ and
the field $\phi$, for some value in the $\lambda_{eff}$ cosmological term 
and noncommutative $\theta$ parameter,
present a dynamical isotropization up to a critical cosmic time $t_{c}$;
 after this time, the effects of  isotropization in the noncommutative minisuperspace   
  seems to disappear. In the quantum regimen, the probability density 
 presents a new structure that corresponds to the value of the noncommutativity parameter.

\end{abstract}

\maketitle

\section{Introduction}
The inclusion of matter into the homogeneous cosmologies is usually
done  with  the inclusion of scalar fields, in order to study different scenarios, such as
inflation, dark matter, dark energy. However, since  the early
seventies, the problem of  the appropriate sources of matter to
solve an  specific scenario \cite{ryan1,ryan}, and its corresponding
lagrangian exists.  In this paper we consider the scalar field and
perfect fluid sources as a first approximation in  the
noncommutative quantization regime,  and in particular for the
Bianchi type I cosmological model, which is the generalization of
the flat FRW cosmological model. 
 Bianchi cosmological models
are the  homogeneous but  anisotropic generalization of the FRW cosmological model, in which the isotropy is not required.
In view of the present data that implies a high degree isotropy in the Universe, the Bianchi model need an
isotropization mechanism to agree with the observation. Otherwise they can only model earlier stages
of the Universe. In the past different kinds o matter have been used to obtain the isotropization. 
In this paper we  explore another possibility to aquire the isotropization in the cosmological model: that 
noncommutativity among the material content (fields and perfect fluid) and the
geometric variables of the cosmological models produces the expected isotropy. We will explore the effects of noncommutativity
in the classical and quantum regimes.

In the last years there has been several attempts to study the
possible effects of noncommutativity in the cosmological scenario.
In particular, in \cite{ncqc}, the authors in a cunning way avoided
the difficult technicalities of analyzing noncommutative
cosmological models, when these are derived from a noncommutative
theory of gravity \cite{ncsdg}. Their proposal is to introduce the
effects of noncommutativity in quantum cosmology, by a deformation
of the minisuperspace, and it is achieved by means of a Moyal
deformation of the WDW equation, similar to the case of the
noncommutative quantum mechanics \cite{gamboa, jabbari}. Some work
has been done in this direction, for example in \cite{barbosa1} the
authors study the implications of noncommutative geometry in
minisuperspace variables for an FRW universe with a conformally
coupled scalar field, using the Bohmian formalism of quantum
trajectories \cite{barbosa2}, also in \cite{pimentel} a
noncommutative deformation with  a scalar field in the scalar-tensor
theory of  gravity was considered. Recently, some work on
noncommutative fields has appeared, see for example Carmona et al.
\cite{carmona}, where the authors gave the generalization to the
noncommutative harmonic oscillator  and have shown the most general
transformation between the coordinates and momenta. In Vakili et al.
\cite{vakili} a  similar procedure is applied to Bianchi I models
without matter, and in \cite{eri} the matter is included. 

 In order to obtein a deformed model, there are  several  ways to proceed, such as \cite{gup}, 
 where the uncentainty relations are generalized, which can be of interest,   in order to obtain new information 
 in the cosmological scenario and other physical quantities. 
 An alternative method that consider deformed special relativity, is considered in \cite{dsr},
where the authors study the effects of the existence of a fundamental length in a cosmological scenario, this is 
described by what is known as the $\kappa$-deformation \cite{kappa}. 
Both approaches have the same role, to obtain fundamental
length, because is generally believe that is a natural feature in all theories which have the duty to obtain the quantization of
gravity.

 The aim of this paper is to construct a noncommutative scenario for
the Bianchi class A cosmological models, introducing noncommutative
scale factors, considering that all minisuperspace  variables $\rm
q^i$ do not commute, so the symplectic structure is modified, in
order to study the consequences of the noncommutative  in  the
universe, such as the isotropization of the anisotropic parameters $\rm \beta_\pm$,
so the decay of the field $\phi$ dynamically.  The other methods could be implemented and
carry the analysis if those methods give a better   dynamical isotropization of the anisotropic parameters.
This will be reported in future work.

The paper is then organized as follows. In Sec. II, we obtain  the
Wheeler-DeWitt equation including several matter contribution, a
scalar field, a  barotropic perfect fluid and a cosmological term.
Also we obtain the corresponding classical solutions.
In Sec. III we present  the commutative quantum solutions for the
cosmological  Bianchi type I, considering various scenarios,
including the effect to the $\rm \Lambda$ cosmological term. In Sec.
IV and V, the corresponding non commutative classical and quantum solutions are
presented, taking that all the  minisuperspace variables do not
commute. Final remarks, are given in Sec. VI.

\section{The Wheeler-DeWitt equation}

Recalling the canonical formulation of the ADM formalism of the
diagonal Bianchi Class A cosmological models. The metrics have the form
\begin{equation}
\rm ds^2= -(N^2- N^j N_j)dt^2 + e^{2\Omega(t)} e^{2\beta_{ij}(t)} \,
\omega^i \omega^j,
\label{metrica}
\end{equation}
where $\rm N$ and $\rm N_i$ are the lapse and shift functions,
respectively, $\Omega(t)$ is a scalar and $\rm \beta_{ij}(t)$ a 3x3
diagonal matrix, $\rm \beta_{ij}=diag(\beta_+ +\sqrt{3}
\beta_-,\beta_+ -\sqrt{3} \beta_-, -2\beta_+)$, $\rm \omega^i$ are
one-forms that  characterize  each cosmological Bianchi type model,
 that obey $\rm d\omega^i= \frac{1}{2} C^i_{jk} \omega^j \wedge
\omega^k,$ $\rm C^i_{jk}$ the structure constants of the
corresponding invariance group \cite{ryan}. The metric for the
Bianchi type I, takes the form
\begin{equation}
\rm ds^2_I=  - N^2dt^2 + e^{2\Omega} e^{2\beta_+ +2\sqrt{3}\beta_-}
dx^2 +e^{2\Omega} e^{2\beta_+ -2\sqrt{3}\beta_-} dy^2 + e^{2\Omega}
e^{-4\beta_+ } dz^2, \label{bi}
\end{equation}

The Lagrangian function is given by
\begin{equation}
\rm L_{Total}=\sqrt{-g}\, (R-2\Lambda)+ L_{matter},
\label{lagraq}
\end{equation}
we use as a first approximation a perfect fluid and a scalar field
as the matter content,  in a comoving frame \cite{ryan,pazos},
\begin{equation}
\rm L_{Total}=\sqrt{-g}\, (R-2\Lambda+ 16\pi G \rho +
\frac{1}{2} g^{\mu\nu}\partial_\mu \phi \partial_\nu \phi + V(\phi) ),
\label{lagra}
\end{equation}
and  using (\ref{bi}), we obtain the corresponding Hamiltonian
function
\begin{equation}
\rm H= \frac{e^{-3\Omega}}{24} \left[-P_\Omega^2 + P_+^2 + P_-^2 +12
P_\phi^2 -48\Lambda  e^{6\Omega} +384\pi G M_\gamma e^{-3(\gamma-1)\Omega} +
12e^{6\Omega}V(\phi) \right], \label{hami1}
\end{equation}
together
with barotropic state equation $\rm p=\gamma \rho$.
The WDW equation for these models is  obtained by the usual identification,
$\rm P_{q^\mu}$
by $\rm -i \partial_{q^\mu}$ in (\ref {hami1}),
with $\rm q^\mu=(\Omega, \beta_+,\beta_-,\phi)$.

Following Hartle and Hawking {\cite {HaHa}}, we consider a
semi-general factor ordering which  gives
\begin{equation}
  \rm  e^{-3\Omega} \left[\frac{\partial^2}{\partial \Omega^2}-
\frac{\partial^2}{\partial \beta_+^2} -
  \frac{\partial^2}{\partial \beta_-^2} -12\frac{\partial^2}{\partial
\phi^2}  +Q\frac{\partial}{\partial \Omega}
  -48\Lambda  e^{6\Omega}+ 384\pi G M_\gamma e^{-3(\gamma-1)\Omega} + 12
e^{6\Omega} V(\phi)\right]\Psi=0 . \label{wdw1}
\end{equation}
This equation is not easy to solve in general, therefore we consider
several particular cases that yield analytical solutions, always
considering the scalar field.

\subsection{Classical Solutions via Hamiltonian Formalism}
To find the commutative equation of motion, we use the
classical phase space variables $\rm q^\mu=(\Omega, \beta_\pm, \phi)$,
where the Poisson algebra for these minisuperspace variables are
\begin{equation}
\rm \left\{ q^\mu, q^\nu  \right\}=0 \qquad \left\{q^\mu,P_{q^\nu} \right\}=\delta^\mu_\nu,
\label{cbracket}
\end{equation}
and recalling the Hamiltonian equation (\ref{hami1}), we obtain the
classical solutions.

The classical equations of motion for the phase variables $q^\mu$ and  $P_{q^\nu}$ are
\begin{eqnarray}
\rm \dot{\Omega}&=&\rm \{\Omega, H\}=-\frac{1}{12} e^{-3\Omega}P_\Omega, \label{do1}\\
\rm \dot{\beta_-}&=&\rm \{\beta_-, H\}=\frac{1}{12} e^{-3\Omega}P_-,\label{beta0-}\\
\rm \dot{\beta_+}&=&\rm \{\beta_+, H\}=\frac{1}{12} e^{-3\Omega} P_+,\label{beta+}\\
\rm \dot{\phi}&=&\rm \{\phi, H\}= e^{-3\Omega} P_\phi,\label{phi0}\\
\rm \dot{P}_\Omega&=&\rm \{P_\Omega, H\}= \frac{1}{8}e^{-3\Omega}
\left[-P^2_\Omega+ P_-^2 + P_+^2 +12P_{\phi}^{2} \right]\nonumber
\\ &&+\left[48\Lambda e^{6\Omega}+\gamma b_\gamma
e^{-3(\gamma-1)\Omega}  -12e^{6\Omega}V(\phi)\right] ,\label{gauge1}\\
\rm \dot{P}_-&=&\rm \{P_-, H\}=0,\qquad \to P_-=\pm p_1 =const. \label{p--}\\
\rm \dot{P}_+&=&\rm \{\phi, H\}=,\qquad \to \qquad  P_+=\pm n_1
=const.\label{p++} \\
\rm \dot{P_{\phi}}&=&\rm \{P_+,
H\}=\frac{1}{8}e^{-3\Omega}\frac{dV(\phi)}{d\phi}. \label{classical}
\end{eqnarray}

In order to have an analytic solution, we have consider different
cases
\begin{equation}
V(\phi)=V_0 =cte. \to P_\phi=const =\frac{m_1}{\sqrt{12}}
\end{equation}
we choose the constant $P_\phi$ in such way for   simplicity.
In this case, we can associate the potential field with the cosmological term, so we have an effective parameter
 $\lambda_{eff}=48\Lambda -12V_{0}$ for $\gamma\not=-1$, and $\lambda_{eff}=48\Lambda-384\pi G M_{-1}-12V_0$ for the case $\gamma=-1$.
We  obtain the relation for $\rm P_\Omega$
\begin{equation}
\rm P_\Omega= \pm \sqrt{a_1^2- \lambda_{eff} e^{6\Omega}+ b_\gamma
e^{-3(\gamma-1) \Omega}}, \label{ppomega}
\end{equation}
where $\rm a_1^2=n_1^2+p_1^2+m_1^2$. Employing this   relation together with Eq.(\ref{do1}),
for  particular epochs of the universe evolution, characterized by different values of the $\gamma$ parameter,
we present the classical solutions in table \ref{tabc1}.

\begin{center}
\begin{tabular}{|l| l|}\hline
 Case & Commutative solutions\\ \hline
$\gamma=-1,\ \ \lambda_{eff} \not=0,\ \ \rm \rho_{-1}=M_{-1}$&$\rm \Omega=  \frac{1}{3} \, Ln \, 
\left[ \frac{e^{2qt}-4a_1^2}{16qe^{qt}} \right]$\\
 $\rm  q^2=-\frac{\lambda_{eff}}{16}$,
 &$\rm \beta_+= \pm\frac{2}{3}\frac{n_1}{a_1} \, arctanh\, \left[\frac{e^{qt}}{2a_1} \right]$\\
 $\rm a_1^2=n_1^2 + p_1^2 +m_1^2,$&$\rm \beta_-= \pm\frac{2}{3}\frac{p_1}{a_1} \, arctanh\, \left[\frac{e^{qt}}{2a_1} \right].$\\
&$\rm \phi= \pm\frac{4}{\sqrt{6}}\frac{m_1}{a_1} \, arctanh\, \left[\frac{e^{qt}}{2a_1} \right],$ 
\\ \hline  $\gamma=1,\ \ \lambda_{eff} <0,\ \ \rm \rho_1=M_1e^{-6\Omega}$
&$\rm \Omega=  \frac{1}{3} \, Ln \, \left[ \frac{e^{2qt}-4a_1^2}{16qe^{qt}} \right],$ \\
$\rm  q= \sqrt{3|\Lambda-\frac{V_0}{4}|},$ &$ \rm \beta_+=\pm  \frac{2}{3}\frac{n_1}{a_1} \, arctanh\, \left[\frac{e^{qt}}{2a_1} \right],$\\
&$ \rm \beta_-=  \pm\frac{2}{3}\frac{p_1}{a_1} \, arctanh\, \left[\frac{e^{qt}}{2a_1} \right].$ \\
$\rm a_1^2=n_1^2 + p_1^2+m_1^2+384\pi G M_1,$
&$ \rm \phi=  \pm\frac{4}{\sqrt{6}}\frac{m_1}{a_1} \, arctanh\, \left[\frac{e^{qt}}{2a_1} \right].$  
\\ \hline
$\gamma=1,\ \ \lambda_{eff} =0,\ \  \rm \rho_1=M_1e^{-6\Omega}$&$\rm \Omega=\frac{1}{3}Ln\, 
[\frac{a_1}{4} t]$, \\
$\rm  a_1^2=n_1^2 + p_1^2+m_1^2 +384\pi G M_1$,&$\rm \beta_+=\pm Ln\,[t^{-\frac{n_1}{3a_1}}],$ \\
&$\rm \beta_-=\pm Ln\,[t^{-\frac{p_1}{3a_1}}].$ \\
&$\rm \phi=\pm Ln\,[t^{-\frac{2m_1}{\sqrt{6}a_1}}].$ \\ \hline
$\gamma=0,\ \ \lambda_{eff} =0,\ \  \rm \rho_0=M_0e^{-3\Omega}$&$\rm \Omega=\frac{1}{3}Ln\, 
\left[ \frac{b_0 t^2}{64} +\frac{a_1 t}{4} \right]$, \\
 $\rm b_0=384 \pi G M_0,$&$\rm \beta_+= \pm \frac{n_1}{3 a_1}\, Ln\,\left[\frac{16a_1 + b_0 t}{t} \right]$, \\
$\rm a_1^2=n_1^2+p_1^2+m_1^2,$&$\rm \beta_-=\pm \frac{p_1}{3 a_1}\, Ln\,\left[\frac{16a_1 + b_0 t}{t} \right].$\\
&$\rm \phi=\pm \frac{4m_1}{\sqrt{6} a_1}\, Ln\,\left[\frac{16a_1 + b_0 t}{t} \right].$\\ \hline
\end{tabular}

\emph{\label{tabc1} Table \ref{tabc1}. Classical Solutions for $\gamma=-1,1,0$,
and constraints $q$, $a_1$ and $b_0.$}
\end{center}
In all cases $\beta_\pm$ are not bounded, excluding isotropization. In the present work we will find out how the 
noncommutativity solve this problem.

\section{Commutative quantum solution}

In the following we consider different cases of the potential  that allowed us to obtain exact solutions.
\subsection{$\rm V(\phi)=V_0$}
For this case, the potential field is equivalent to a cosmological
term, and  it is included in the $\rm \Lambda_{eff}$ parameter. Using
the method of separation of variables, 

$ \rm \Psi(\Omega,\beta_\pm,\phi)= {\cal A}(\Omega){\cal
B}(\beta_+){\cal C}(\beta_-){\cal D}(\phi), $ in  Eq. (\ref{wdw1})
we find
\begin{eqnarray}
\rm \frac{d^2 {\cal A}}{d\Omega^2} +  Q \frac{d{\cal A}}{d\Omega} +
\left( 384 \pi GM_\gamma e^{-3(\gamma-1) \Omega} +
 \Lambda_{eff} e^{6\Omega}  +\alpha^2 \right) {\cal A} &=& 0,
\label{omega}\\
\rm  \frac{d^2 {\cal B}}{d\beta_+^2} +\nu^2 {\cal B} &=& 0, \label{beta--}\\
\rm  \frac{d^2 {\cal C}}{d\beta_-^2} + \mu^2 {\cal C} &=& 0, \label{beta++}\\
\rm \frac{d^2 {\cal D}}{d\phi^2}+ \eta^2 {\cal D}&=&0, \label{phi}
\end{eqnarray}
where $\rm \Lambda_{eff}=+12V_0 - 48 \Lambda$ is the effective
cosmological constant now, and the separation constants  are related
by $\mu^2=\alpha^2-\nu^2-12\eta^2$. The solutions to (\ref{beta--}, \ref{beta++})
and (\ref{phi}) are
\begin{eqnarray}
\rm {\cal B}&=& \rm A_1\, e^{i\nu \beta_+} +B_1\, e^{-i\nu \beta_+},\label{++} \\
\rm {\cal C}&=& \rm A_2\, e^{i\mu \beta_-} +B_2\, e^{-i\mu \beta_-}, \label{--}\\
\rm {\cal D}&=& \rm A_3\,  e^{i\eta \phi} +B_3\, e^{-i\eta \phi},\label{pp}
\end{eqnarray}
where  $\rm A_j, B_j$, j=1,2,3 are integration constants.

The general solution to  equation  (\ref{omega}) is very complicated
to find, so in the following we will solve
particular for particular cases.\\

\subsubsection{ $\rm \Lambda_{eff}=0$,  $\gamma\not=1$, and factor ordering
$\rm Q\not=0$.}

For this choice of the parameters, the equation  (\ref{omega}) is
\begin{equation}
\rm \frac{d^2 {\cal A}}{d\Omega^2} +  Q \frac{d{\cal A}}{d\Omega}
+ \left( 384 \pi GM_\gamma e^{-3(\gamma-1) \Omega}+ \alpha^2\right){\cal A}
= 0 , \label{00}
\end{equation}
this equation can be written as an ordinary Bessel equation by making the
transformation
$\rm z=z_0 e^{-3(\gamma-1) \Omega}, ~ z_0= 384\pi G M_\gamma$,
\begin{equation}
\rm z^2 \frac{d^2{\cal A}}{dz^2} + \left(1+\frac{Q}{9(\gamma-1)^2}\right)
z\frac{d{\cal A}}{dz} +\frac{1}{9(\gamma-1)^2} \left(z+ \alpha^2 \right){\cal A}=0,
\end{equation}
whose  solution becomes
\begin{equation}
\rm {\cal A}_\rho=  z^{q_0} J_{\pm
i\rho}\left(\frac{2}{3(\gamma-1)}z^{\frac{1}{2}}\right), \qquad
q_0=-\frac{Q}{18(\gamma-1)^2},
\end{equation}
where $\rm \rho=\frac{1}{3(\gamma-1)} \sqrt{4\alpha^2
-\left(\frac{Q}{3(\gamma-1)}\right)^2}$,
with $\rm \pm i\rho$  the  order of
the ordinary Bessel function.

Then, we  find  For the complete  wave function, we find the equation
\begin{eqnarray}
&&\rm \Psi(\Omega,\beta_\pm,\phi)=\rm z^{q_0} J_{\pm i\rho} \left( \frac{2 \; z^{\frac{1}{2}} }{3(\gamma-1)}\right)
\left[A_3\,  e^{i\eta \phi} +B_3\, e^{-i\eta \phi}\right]\nonumber\\
&&\rm \left( A_1\, e^{i\nu \beta_+} +B_1\, e^{-i\nu \beta_+}\right)
\left[ A_2\, e^{i\mu \beta_-} +B_2\, e^{-i\mu \beta_-}\right].
\label{one}
\end{eqnarray}

To obtain the complete  wavepacket out of the wavefunction is  complicated, that is because all parameters are interrelated, 
so  we limit the study   to build  3D wavepacket by considering two variables field. 
This is carry out for the case where the parameter  $\gamma$  take the value 
-1 (inflation phenomenon).

\subsubsection{$\rm \Lambda_{eff}\ne0$,  $\gamma=1$, factor ordering $\rm
Q\not=0$.}

When we introduce the effective cosmological constant $\rm
\Lambda_{eff}$ and a stiff fluid,  equation  (\ref{omega}) takes the
following form
\begin{equation}
\rm \frac{d^2 {\cal A}}{d\Omega^2} + Q \frac{d{\cal A}}{d\Omega}
+ \left( \Lambda_{eff} e^{6\Omega}+ c_1 \right){\cal A} = 0 ,
\label{0000}
\end{equation}
where $\rm c_1= 384 \pi GM_1 + \alpha^2$,
and making the transformation $\rm z=\Lambda_{eff} e^{6\Omega}$, we
find the solutions in terms of generic Bessel functions
\begin{eqnarray}
\rm {\cal A}_\rho (z)&=&\rm  z^{-\frac{Q}{12}} \, Z_\rho \left(\pm\frac{1}{3}z^{\frac{1}{2}}\right), \\
\rm {\cal A_\rho}(\Omega)&=& \rm A_0 \, e^{-\frac{Q}{2}\Omega}\, Z_\rho \left(
\pm \frac{1}{3} \sqrt{\lambda} e^{3\Omega}\right). \label{four}
\end{eqnarray}
with order $\rm \rho=\frac{1}{6}\sqrt{Q^2 -4 c_1}$ and $\rm Z_\rho$ is a generic Bessel function.

\subsubsection{ $\rm \Lambda_{eff}\not=0$,  $\gamma=-1$, factor ordering
$\rm Q\not=0$}

The other interesting case, corresponds to $\rm \gamma=-1$ (the
equation of state for inflation), proceeding as before, we make the
transformation $\rm z=b e^{6\Omega}$, where $\rm b=384\pi G M_{-1}
+\Lambda$, we arrive to equation
\begin{equation}
\rm z^2 \frac{d^2{\cal A}}{dz^2} + \left(1+\frac{Q}{6}\right) z\frac{d{\cal
A}}{dz} +
\frac{1}{36}\left(z+\alpha^2\right){\cal A}=0,
\end{equation}
the solution is written with the generic Bessel functions
\begin{equation}
\rm {\cal A}(z)=  z^{-\frac{Q}{12}} \, Z_{\pm i\rho} \left(\pm
\frac{1}{3} z^{\frac{1}{2}}\right),
\end{equation}
with order $\rm \rho=\frac{1}{6}\sqrt{4\alpha^2 -Q^2}$.

\begin{figure}
\includegraphics[width=12cm]{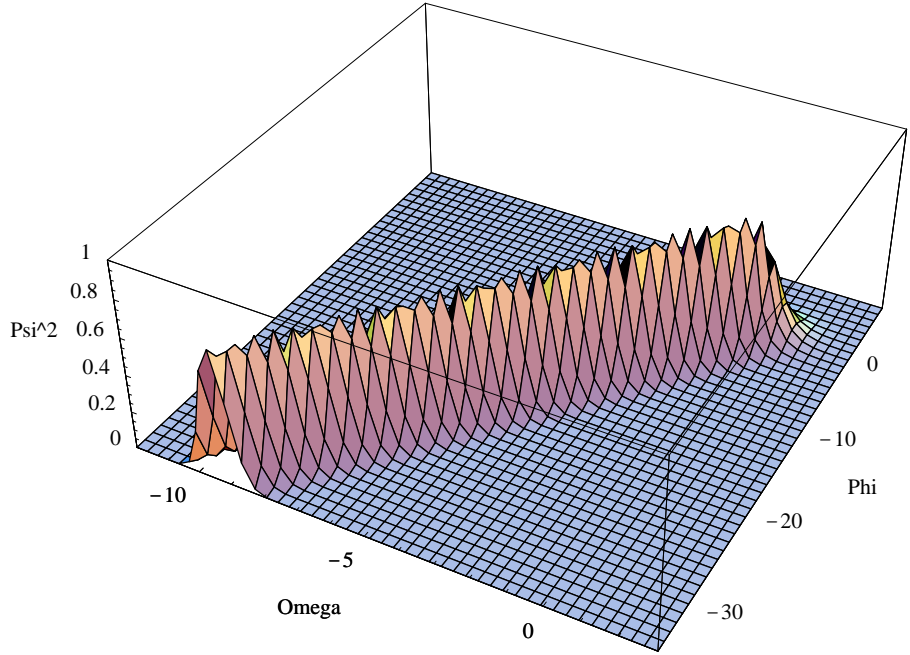}
\caption{\label{phi-gra}  3D plot$\rm |\Psi|^2$ corresponding to the Gaussian wavepacket 
equation (\ref{w-p}) in the commuting case, with $\mu=\nu=0$ }
\end{figure}

From equations (\ref{++}, \ref{--}, \ref{pp}) we notice that the dependence of the wavefunction on the variables
$\phi$ and $\beta_\pm$ have the same structure. In the aim of building a Gaussian weighted wavepacket \cite{ki,gra},
we fix the attention to  solutions in two of the four variables. For simplicity we choose $\Omega$ and the scalar field $\phi$. 
We also choose the particular 
case in which the  factor ordering $Q$ is zero, 
\begin{equation}
\rm \Psi(\Omega,\phi)={\cal N}\int_{-\infty}^\infty e^{-a(\rho-b)^2}\, e^{-\frac{\sqrt{3}}{2}\rho \phi}\,J_{i\rho}\left(\frac{1}{3}\sqrt{b}
e^{3\Omega} \right) d\rho,
\label{w-p}
\end{equation}
the corresponding plot is presented in figure (\ref{phi-gra}), taking the particular values for a=1.5 and b=1.3; 
and as in the \cite{ncqc,gup} the integral was performed numerically. In this plot the wave function  
is not peaked around any particular values of the variables $(\Omega,\phi)$.

\subsection{$\rm V(\phi)=V_0e^{- \beta\phi}$, stiff fluid $\gamma=1$ and
$\Lambda=0$}

The equation (\ref{wdw1}) becomes
\begin{equation}
  \rm   \left[\frac{\partial^2}{\partial \Omega^2}-
\frac{\partial^2}{\partial \beta_+^2} -
  \frac{\partial^2}{\partial \beta_-^2} -12\frac{\partial^2}{\partial
\phi^2}  +Q\frac{\partial}{\partial \Omega}
  + 384\pi G M_1 + 12 V_0e^{6\Omega-\beta\phi} \right]\Psi=0
\label{wdw2}
\end{equation}
using the ansatz,
$
\rm \Psi(\Omega,\beta_\pm,\phi)= {\cal F}(\Omega,\phi){\cal B}(\beta_+){\cal
C}(\beta_-),
$
on Eq. (\ref{wdw2}) we find
\begin{eqnarray}
\rm \frac{1}{{\cal F}}\frac{\partial^2 {\cal F}}{\partial \Omega^2} +
\frac{Q}{{\cal F}}
\frac{\partial {\cal F}}{\partial \Omega} &-&
\rm  \frac{12}{{\cal F}}  \frac{\partial^2 {\cal F}}{\partial \phi^2}   +
     12 V_0e^{6\Omega-\beta \phi}  +c   = 0, \label{omega-phi}\\
\rm  \frac{d^2 {\cal B}}{d\beta_+^2} +\nu^2 {\cal B} &=& 0, ~~ \rm
\frac{d^2 {\cal C}}{d\beta_-^2} + \mu^2 {\cal C} = 0, \label{beta-}
\end{eqnarray}
where $\rm c=\alpha^2 + 384 \pi GM_1$, and the constants satisfy the
relation $\rm \mu^2=\alpha^2 - \nu^2$.

There are many changes of variables for   equation (\ref{omega-phi}) that
allows to solve it by separation of variables, we choose the
following one
\begin{equation}
\rm x=-6\Omega + \beta \phi, \qquad y= -2\Omega + \frac{\phi}{\beta},
\label{xy}
\end{equation}
which is valid for $\beta^2 \ne 3$,  then  equation
(\ref{omega-phi}), using ${\cal F}= X(x) Y(y)$, is transformed into the
following two equations
\begin{eqnarray}
&& \rm\frac{d^2 X}{dx^2}-\frac{Q}{2(3-\beta^2)}\frac{dX}{dx} + 
+\frac{1}{12(3-\beta^2)}\left[12V_0 e^{-x} +n^2\right]X=0, \label{x}\\
&& \rm \frac{d^2
Y}{dy^2}+\frac{2Q\beta^2}{4(3-\beta^2)}\frac{dY}{dy}-\frac{\beta^2
m^2}{4(3-\beta^2)}Y =0,\label{y}
\end{eqnarray}
with $\rm n^2=c-m^2$. Substituting $\rm z=e^{-x}$ in (\ref{x}), we
obtain a Bessel differential equation whose solution is 
\begin{equation}
\rm X(x)=Exp[-\frac{Q}{4(3-\beta^2)}x]J_{\pm r}\left(
\sqrt{\frac{V_0}{3-\beta^2}} e^{-x/2} \right),
\end{equation}
with $\rm r=\sqrt{\frac{Q^2}{4(3-\beta^2)^2}-\frac{n^2}{3(3-\beta^2)}}$, the
corresponding order.
The other equation has the solution
\begin{equation}
\rm Y(y)= e^{sy}\left[A_4 e^{\frac{Q\beta^2}{4(3-\beta^2)}y}+ B_4
e^{-\frac{Q\beta^2}{4(3-\beta^2)}y} \right],
\end{equation}
with $\rm s=\sqrt{\left(\frac{Q\beta^2}{4(3-\beta^2)} \right)^2 +
\frac{m^2 \beta^2}{4(3-\beta^2)}}$.

For the remaining case, $\beta^2 = 3$ the solution to the
differential equation is
\begin{equation}
{\cal F}=F_1 e^{-\frac{Q\Omega}{2}}[e^{\frac{6b x +a e^{6x}}{6
\kappa}} e^{-\frac{\kappa y}{4}}],
\end{equation}
here $\kappa$ is a non null separation constant, $F_1$ is an
arbitrary integration constant, $a=c-Q^2/4$, $b= 12 V_0$,  and the
new variables $x$ and $y$ are
\begin{equation}
x= \frac{6 \Omega \mp \sqrt{3}\phi}{6}, y= \frac{6 \Omega \pm
\sqrt{3}\phi}{6}.\label{tranf}
\end{equation}.

\section{Noncommutative classical solution}
 In the commutative model we know that the solutions to Hamilton's equations are the same as 
 in General Relativity. Now the natural extension is to consider the noncommutative version
of our model, with the idea of noncommutative between the four
variables $\rm (\Omega_{nc}, \beta_{\pm nc},\phi)$, so we apply a deformation of the Poisson
algebra. For this we start with the usual Hamiltonian (\ref{hami1}), but the symplectic
structure is modify as follow
\begin{eqnarray}
\rm \left\{P_{\Omega},P_{\pm } \right\}_\star&=&\rm\left\{P_{+},P_{-}
\right\}_\star=0, \qquad \left\{q^\mu,P_{q^\mu} \right\}_\star=1, \\
\rm \left\{\Omega, \beta_+ \right\}_\star&=&\rm i\theta_1, \quad \left\{\Omega,\beta_- \right\}_\star= i\theta_2,
\quad \left\{\beta_-,\beta_+ \right\}_\star= i\theta_3,\nonumber\\
\rm \left\{\Omega,\phi\right\}_\star&=&\rm i\theta_4, \quad \left\{\phi,\beta_+\right\}_\star=i\theta_5, \quad
\left\{\phi,\beta_-\right\}_\star=i\theta_6.
\label{ncbracket}
\end{eqnarray}
where the $\star$ is the Moyal product \cite{Szabo2}, and  the Hamiltonian
 resulting is
\begin{equation}
\rm H_{nc}= \frac{Ne^{-3 \Omega_{nc}}}{24} \left[-P_{\Omega}^2 + P_{+}^2 + P_{-}^2+12P_\phi^2 -\lambda  e^{6 \Omega_{nc}}
  +12V(\phi) e^{6 \Omega_{nc}}  +b_\gamma e^{-3(\gamma-1) \Omega_{nc}} \right]=0,
 \label{non-hami11}
 \end{equation}
but the symplectic structure is the one that we know, the
commutative one (\ref{cbracket}).
There are two formalism to study the  noncommutative equations of motion, for the first formalism that
we exposed have  the original variables, but with the modified
symplectic structure,
\begin{eqnarray}
\dot{q^\mu_{nc}}&=&\rm \{q^\mu, H\}_\star ,
\nonumber\\
\dot{P^\mu_{nc}}&=&\rm \{P^\mu, H\}_\star ,
\end{eqnarray} and for the second formalism we use the
shifted variables but with the original (commutative) symplectic
structure
\begin{eqnarray}
\dot{q^\mu_{nc}}&=&\rm \{q^\mu_{nc}, H_{nc}\} ,
\nonumber\\
\dot{P^\mu_{nc}}&=&\rm \{P^\mu_{nc}, H_{nc}\} ,
\end{eqnarray}
in  both approaches we have the same result. Therefore the equations of motion take the form
\begin{eqnarray}
\rm {\dot \Omega}_{nc}&=&\rm \{\Omega, H\}_\star=\{\Omega_{nc},
H_{nc}\}=-\frac{e^{-3 \Omega_{nc}}}{12} P_{\Omega},\\
\rm {\dot \beta}_{-nc}&=&\rm \{\beta_{-}, H\}_\star=\rm \{\beta_{-nc}, H_{nc}\}=\frac{e^{-3 \Omega_{nc}}}{12}P_{-}+\frac{\theta_2 }{2} \dot P_{\Omega}
-\frac{\theta_6}{2}P_\phi   ,\\
\rm {\dot \beta}_{+nc}&=& \rm \{\beta_{+}, H\}_\star=\{\beta_{+nc}, H_{nc}\}=\frac{e^{-3 \Omega_{nc}}}{12} P_{+} +\frac{\theta_1}{2} \dot P_{\Omega}
-\frac{\theta_5}{2} P_\phi ,\label{nc22} \\
\rm {\dot \phi}_{nc}&=&\rm \rm \{\phi, H\}_\star=\{\phi_{+nc}, H_{nc}\}=e^{-3\Omega_{nc}}P_\phi-\frac{\theta_4}{2}P_\Omega+\frac{\theta_5}{2}P_+
+\frac{\theta_6}{2}P_- \label{phinc}\\
\rm \dot{P}_\Omega&=&\rm \{P_{\Omega},H\}_\star=\rm
\{P_\Omega,H_{nc}\}=\frac{e^{-3 \Omega_{nc}}}{8}\left[6\lambda e^{6 \Omega_{nc}}+3(\gamma-1)
b_\gamma e^{-3(\gamma-1) \Omega_{nc}} \right],\label{ppomegac}\\
\rm \dot{P}_-&=&\rm \{P_{-}, H\}_\star=\rm \{P_-, H_{nc}\}=0, \quad \to \quad P_-=p_1,\label{nc3} \\
\rm \dot{P}_+&=&\rm \{P_+, H\}_\star=\rm \{P_+,H_{nc}\}=0, \quad \to \quad P_+=n_1. \label{nc4}\\
\rm \dot{P}_\phi&=&\rm \{P_\phi, H\}_\star=\rm \{P_\phi,H_{nc}\}=0, \quad \to \quad P_\phi=\frac{m_1}{\sqrt{12}}. \label{nc5}
\label{classicalnc}
\end{eqnarray}
if we proceed as in the commutative case we get the solutions showed in the table \ref{tabnc1}.

\begin{center}
\begin{tabular}{|l| l|}\hline
 Case & Noncommutative Solutions\\ \hline
$\gamma=-1,\ \ \lambda_{eff} \not=0,\ \ \rm \rho_{-1}=M_{-1}$&$\rm \Omega_{nc}=\frac{1}{3} \, Ln \, \left[ \frac{e^{2qt}-4a_1^2}{16qe^{qt}}\right]
-\frac{\theta_1}{2}p_1-\frac{\theta_2}{2} n_1+\frac{\theta_4}{2\sqrt{12}}m_1$,  \\
$\rm a_1^2=n_1^2 + p_1^2+m_1^2$ ,&$\rm \beta_{+nc}= \pm\frac{2}{3}\frac{n_1}{a_1} \, arctanh\, \left[\frac{e^{qt}}{2a_1} \right]+
\frac{\theta_1}{8}\left(\frac{e^{qt}}{4}+ a_1^2 e^{-q t} \right) -\frac{\theta_3}{2}p_1 -\frac{\theta_6}{2\sqrt{12}}m_1 , $\\
$\rm q^2=-\frac{\lambda_{eff}}{16}$,
&$\rm \beta_{-nc}= \pm\frac{2}{3}\frac{p_1}{a_1} \, arctanh\, \left[\frac{e^{qt}}{2a_1} \right]+
\frac{\theta_2}{8} \left(\frac{e^{qt}}{4}+ a_1^2 e^{-q t} \right)+\frac{\theta_3}{2}n_1-\frac{\theta_5}{2\sqrt{12}}m_1,$ \\
&$\rm \phi_{nc}= \pm\frac{4}{\sqrt{6}}\frac{m_1}{a_1} \, arctanh\, \left[\frac{e^{qt}}{2a_1} \right]-\frac{\theta_4}{2}\left(\frac{e^{qt}}{4}+ a_1^2 e^{-q t} \right)
+\frac{\theta_5}{2}n_1+\frac{\theta_6}{2}p_1,$\\ \hline   
$\gamma=1,\ \ \lambda_{eff} <0,\ \ \rm \rho_1=M_1e^{-6\Omega}$ &
$\rm \Omega_{nc}=  \frac{1}{3} \, Ln \, \left[\frac{e^{2qt}-4a_1^2}{16qe^{qt}} \right]-\frac{\theta_1}{2}p_1
-\frac{\theta_2}{2} n_1+\frac{\theta_4}{2}\frac{m_1}{\sqrt{12}},$ \\
 $\rm a_1^2=n_1^2 + p_1^2+m_1^2+384\pi G M_1,$ &$ \rm \beta_{+nc}=\pm  \frac{2}{3}\frac{n_1}{a_1} \, arctanh\,
\left[\frac{e^{qt}}{2a_1} \right]+ \frac{\theta_1}{8}
\left(\frac{e^{qt}}{4}+ a_1^2 e^{-q t} \right)-\frac{\theta_3}{2}p_1-\frac{\theta_6}{2\sqrt{12}} m_1,$\\
 $\rm q= \sqrt{3|\Lambda-\frac{V_0}{4}|},$&$ \rm \beta_{-nc}=  \pm\frac{2}{3}\frac{p_1}{a_1} \, arctanh\, \left[\frac{e^{qt}}{2a_1} \right]
 + \frac{\theta_2}{8}\left(\frac{e^{qt}}{4}+ a_1^2 e^{-q t} \right)+\frac{\theta_3}{2}n_1-\frac{\theta_5}{2\sqrt{12}}m_1,$  \\ 
&$ \rm \phi_{nc}=  \pm\frac{4}{\sqrt{6}}\frac{m_1}{a_1} \, arctanh\, \left[\frac{e^{qt}}{2a_1} \right]
-\frac{\theta_4}{2}\left(\frac{e^{qt}}{4}+ a_1^2 e^{-q t} \right)+\frac{\theta_5}{2}n_1+\frac{\theta_6}{2}p_1.$\\  \hline 
$\gamma=1,\ \ \lambda_{eff} =0,\ \  \rm \rho_1=M_1e^{-6\Omega}$&$\rm \Omega_{nc}=\frac{1}{3}Ln\, [\frac{a_1}{4} t]-\frac{\theta_1}{2}
p_1-\frac{\theta_2}{2} n_1+\frac{\theta_4}{2\sqrt{12}} m_1$, \\
$\rm a_1^2=n_1^2 + p_1^2 +m_1^2+384\pi G M_1$,&$\rm \beta_{+nc}=\pm Ln\,[t^{-\frac{n_1}{3a_1}}]
+ \frac{\theta_1}{2} a_1-\frac{\theta_3}{2}p_1 -\frac{\theta_6}{2\sqrt{12}} m_1,$ \\
&$\rm \beta_{-nc}=\pm Ln\,[t^{-\frac{p_1}{3a_1}}]+ \frac{\theta_2}{2} a_1+\frac{\theta_3}{2}n_1-\frac{\theta_5}{2\sqrt{12}} m_1,$ \\
&$\rm \phi=\pm Ln\,[t^{-\frac{2m_1}{\sqrt{6}a_1}}]-\frac{\theta_4}{2} a_1+\frac{\theta_5}{2} n_1 + \frac{\theta_6}{2} p_1.$ \\ \hline 
$\gamma=0,\ \ \lambda_{eff} =0,\ \  \rm \rho_0=M_0e^{-3\Omega}$&$\rm \Omega_{nc}=\frac{1}{3}Ln\, 
\left[ \frac{b_0 t^2}{64} +\frac{a_1 t}{4}\right]-\frac{\theta_1}{2} p_1-\frac{\theta_2}{2} n_1 +\frac{\theta_4}{2\sqrt{12}} m_1$, \\
$\rm b_0=384 \pi G M_0,$&$\rm \beta_{+nc}= \pm \frac{n_1}{3 a_1}\, Ln\,\left[\frac{16a_1 + b_0 t}{t} \right]
+ \frac{\theta_1}{2} \sqrt{a_1^2 + \frac{b_0 t^2}{64} +\frac{a_1 t}{4}}-\frac{\theta_3}{2}p_1-\frac{\theta_6}{2\sqrt{12}} m_1$,\\
$\rm a_1^2=n_1^2+p_1^2+m_1^2,$&$\rm \beta_{-nc}=\pm \frac{p_1}{3 a_1}\, Ln\,\left[\frac{16a_1 + b_0 t}{t} \right]+
\frac{\theta_2}{2} \sqrt{a_1^2 + \frac{b_0 t^2}{64} +\frac{a_1 t}{4}}+\frac{\theta_3}{2}n_1-\frac{\theta_5}{2\sqrt{12}} m_1.$\\
&$\rm \phi=\pm \frac{4m_1}{\sqrt{6} a_1}\, Ln\,\left[\frac{16a_1 + b_0 t}{t} \right]
-\frac{\theta_4}{2} \sqrt{a_1^2 + \frac{b_0 t^2}{64} +\frac{a_1 t}{4}} +\frac{\theta_5}{2} n_1 + \frac{\theta_6}{2} p_1.$\\ \hline
\end{tabular}\\
\emph{\label{tabnc1} Table \ref{tabnc1}. Noncommutative solutions for,
$\gamma=-1,1,0,$  and constraints $q$, $a_1$
and $b_0$.}
\end{center}

\begin{figure}
\includegraphics[width=12cm]{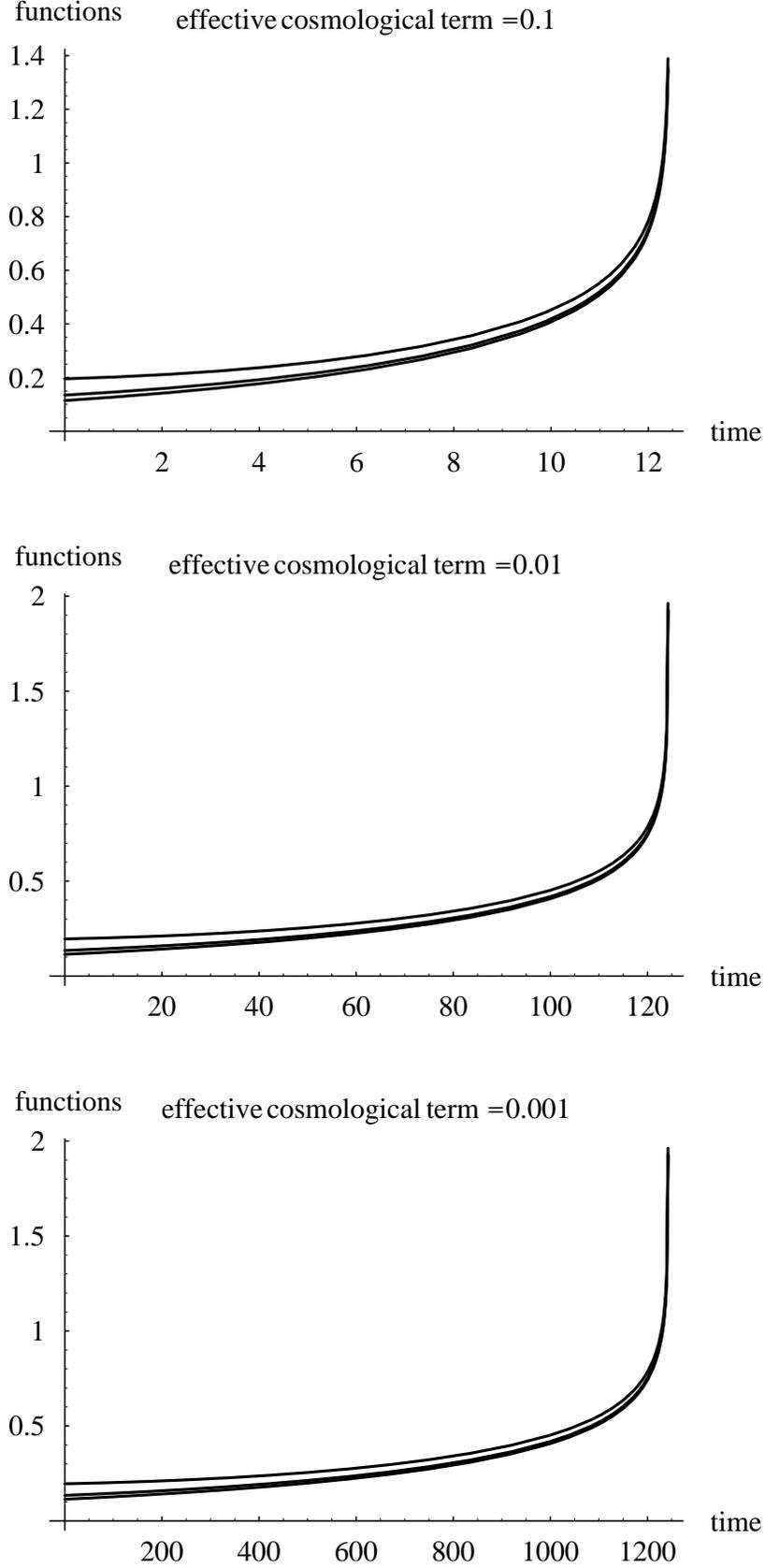}
\caption{\label{be1} Plots of $ \rm \beta_{\pm nc}$ and $\phi$ field, that appear in the second line
in the table \ref{tabnc1},
 using the values in the parameters $\rm n_1=1, p_1=1, m_1=1$ and
$\theta=0,  0.05 , 0.2$, from bottom to top in the figure.
The possible isotropization in the parameters $\beta_{\pm nc}$ and the decay in the $ \phi_{nc}$ field is saw in function of the curvature, 
but it appears again in this fields after critical cosmic time $\rm t_c$ }
\end{figure}
 We see that the noncommutativity parameters $\theta_1$ and $\theta_2$ could be chosen, with same fine tunning, in order to lower the anisotropy at
 the present time, but in the long term the anisotripy   will reappear. The evolution of the isotropy (anisotropy) can be  appreciate in   
 figure (\ref{be1}).
\section{Noncommutative quantum solution}

As already mentioned, we are looking for the noncommuative
deformation of the Bianchi I quantum cosmological model. In order to
find the noncommutative generalization we need to solve the
noncommutative Einstein equation, this is a formidable task, due to
the highly non linear character of the theory \cite{ncsdg},
fortunately we can circumvent these difficulties by following Ref.
\cite{ncqc}. Now  we  can proceed to the  non commutative model,
actually  we will consider, that the minisuperspace  variables $\rm
q^i=(\Omega,\beta_+,\beta_-,\phi)$ do not commute, so that the symplectic
structure is modified as follows
\begin{equation}
\rm [q^i, q^j ]= i\theta^{ij}, \quad [P_i, P_j]=0, \quad
[q^i,P_j]=i\delta^i_j,
\label{rules}
\end{equation}
in particular, we choose the following representation
\begin{eqnarray}
\rm [\Omega, \beta_+ ]&=&\rm i\theta_1, \quad [\Omega,\beta_- ]= i\theta_2,
\quad [\beta_-,\beta_+ ]= i\theta_3,\nonumber\\
\rm [\Omega,\phi]&=&\rm i\theta_4, \quad [\phi,\beta_+]=i\theta_5, \quad
[\phi,\beta_-]=i\theta_6.
\label{rules1}
\end{eqnarray}
where the $\rm \theta_i$ parameters are a measure of the non commutativity
between the minisuperspace variables.
The commutation relation (\ref{rules}) or (\ref{rules1}) are not the most
general deformations that  define a noncommutative field;
 but this relations give us the isotropization of the models that we are working.

It is well known, that this non-commutativity can be formulated in
term of non-commutative minisuperspace functions with the Moyal star
product $\star$ of functions. 
The commutation relations (\ref{rules}) can be implemented in terms of the commuting 
 coordinates of the standard quantum mechanics ( Bopp shift) and it results in a 
 modification of the potential  term of
the WDW equation \cite{ncqc,pimentel},
and one possibility  is, for example,
\begin{eqnarray}
\rm \Omega_{nc} &\to&\rm \Omega + \frac{\theta_1}{2}P_+ +\frac{\theta_2}{2}
P_-+\frac{\theta_4}{2}P_\phi, \nonumber\\
\rm \beta_{-nc}&\to&\rm  \beta_- -\frac{\theta_2}{2}
P_\Omega+\frac{\theta_3}{2}P_+-\frac{\theta_6}{2}P_\phi, \nonumber\\
\rm \beta_{+nc} &\to&\rm  \beta_+-\frac{\theta_1}{2} P_\Omega -
\frac{\theta_3}{2}P_--\frac{\theta_5}{2}P_\phi,\nonumber\\
\rm \phi_{nc}&\to&\rm \phi-\frac{\theta_4}{2}P_\Omega
+\frac{\theta_5}{2}P_++\frac{\theta_6}{2}P_-.
\label{transformation}
\end{eqnarray}
These transformation are not the most general in order to define
noncommutative fields, we can see for example Carmona et al.
\cite{carmona}, where the authors give the generalization of the
noncommutative harmonic oscillator  and have shown the most linear
transformation between the coordinates and momenta. In Vakili et al.
\cite{vakili} a
 similar procedure is applied to Bianchi I models without matter.

However, these shifts  modify the potential term as follows
\begin{eqnarray}
&& \rm U(\Omega,\beta_\pm,\phi,\theta_i) =-48 \Lambda e^{6[\Omega +
\frac{\theta_1}{2}P_+ +\frac{\theta_2}{2} P_-+\frac{\theta_4}{2}P_\phi]}
 +384\pi GM_\gamma e^{-3(\gamma-1)[\Omega + \frac{\theta_1}{2}P_+
+\frac{\theta_2}{2} P_-+\frac{\theta_4}{2}P_\phi]}\nonumber\\
&&\rm +12 e^{6\left[\Omega + \frac{\theta_1}{2}P_+ +\frac{\theta_2}{2}
P_-+\frac{\theta_4}{2}P_\phi\right] } 
V\left(\phi-\frac{\theta_4}{2}P_\Omega
+\frac{\theta_5}{2}P_++\frac{\theta_6}{2}P_-\right).
\end{eqnarray}

Using the generalized Baker-Campbell-Hausdorff formula
\cite{wil},
\begin{equation}
\rm e^{\eta(\hat A + \hat B)}= e^{-\eta^2[\hat A , \hat B]}\, e^{\eta \hat
A} \, e^{\eta  \hat B}
\end{equation}
and the relation between the variables (\ref{rules1}),  we find the
noncommutative WDW (NCWDW) equation
\begin{eqnarray}
  \rm && \left[\frac{\partial^2}{\partial \Omega^2}-
\frac{\partial^2}{\partial \beta_+^2} -
  \frac{\partial^2}{\partial \beta_-^2} -12\frac{\partial^2}{\partial
\phi^2}
  +Q\frac{\partial}{\partial \Omega} 
 -48 \Lambda e^{6[\Omega - \frac{i\theta_1}{2}\partial_+
-\frac{i\theta_2}{2} \partial_- -\frac{i\theta_4}{2}\partial_\phi]}
\right. \nonumber\\
&&\rm \left. +384\pi GM_\gamma e^{-3(\gamma-1)[\Omega -
\frac{i\theta_1}{2}\partial_+ -\frac{i\theta_2}{2} \partial_-
-\frac{i\theta_4}{2}\partial_\phi]}
 +12 e^{6\left[\Omega - \frac{i\theta_1}{2}\partial_+
-\frac{i\theta_2}{2} \partial_-
-\frac{i\theta_4}{2}\partial_\phi\right] } \times
\right. \nonumber\\
&& \rm \left. V\left(\phi+\frac{i\theta_4}{2}\partial_\Omega
-\frac{i\theta_5}{2}\partial_+- \frac{i\theta_6}{2}\partial_-\right).
\right]\Psi=0
\label{wdwnc}
\end{eqnarray}
To solve the equation we are going to focus our efforts on
the same cases that were considered in the commutative model.\\

Lets start with case 1, $\rm \Lambda_{eff}=0$,  $\gamma\not=1$, and factor
ordering $\rm Q\not=0$. We propose an
ansatz of the form
\begin{equation}
\rm \Psi(\Omega,\beta_{\pm},\phi)\equiv  e^{\pm i\eta \phi}e^{\pm
 i\nu\beta_+} e^{\pm i\mu\beta_- }{\cal A}(\Omega) ,
\end{equation}
and taking in account that
$\rm e^{i\theta \frac{\partial}{\partial x}} e^{\eta x}\equiv e^{i\eta
\theta}e^{\eta x}$,
then equation (\ref{wdwnc}) is written as

\begin{equation}
\rm \frac{d^2 {\cal A}}{d\Omega^2} +  Q \frac{d{\cal A}}{d\Omega} +\left(
\alpha^2+       384 \pi GM_\gamma
e^{-3(\gamma-1)\Omega}e^{\mp\frac{3}{2}(\gamma-1)(\theta_4
\eta+\theta_1 \nu+\theta_2 \mu} \right) {\cal A} = 0, \label{omeganc}
\end{equation}
where the separation constants have the following relation between
them, $\mu^2=\alpha^2-\nu^2-12\eta^2$. This equation can be written
as an ordinary Bessel equation by making the transformation $\rm
z=z_0 e^{-3(\gamma-1) \Omega}$, with $\rm z_0= 384\pi G M_\gamma
e^{\mp\frac{3}{2}(\gamma-1)(\theta_4 \eta +\theta_1 \nu+\theta_2
\mu)} $,
\begin{equation}
\rm z^2 \frac{d^2{\cal A}}{dz^2} + \left(1+\frac{Q}{9(\gamma-1)^2}\right)
z\frac{d{\cal A}}{dz}  
 +\frac{1}{9(\gamma-1)^2} \left(z+ \alpha^2 \right){\cal A}=0,
\end{equation}
whose  solution is
\begin{equation}
\rm {\cal A}_\rho=  z^{q_0} J_{\pm
i\rho}\left(\frac{2}{3(\gamma-1)}z^{\frac{1}{2}}\right),
\qquad  q_0=-\frac{Q}{18(\gamma-1)^2}
\end{equation}
with order $\rm \rho=\frac{1}{3(\gamma-1)} \sqrt{4\alpha^2
-\left(\frac{Q}{3(\gamma-1)}\right)^2}$. As expected, in the limit
$\theta_i \to 0$ we get the commutative solutions, so the transition
from noncommutative minisuperspace is well defined.

For case 2, $\rm \Lambda_{eff}\ne0$,  $\gamma=1$, and factor
ordering $\rm Q\not=0$, the NCWDW equation has the ``noncommutative
potential" $\rm U(\Omega,\Lambda_{eff},\theta) =
\left(\Lambda_{eff}e^{6[\Omega - \frac{i\theta_1}{2}\partial_+
-\frac{i\theta_2}{2} \partial_-
-\frac{i\theta_4}{2}\partial_\phi]}+c_1 \right) $. After separating
with $\rm \Psi\equiv  e^{\pm i\eta \phi}e^{\pm i\nu\beta_+} e^{\pm i\mu\beta_- }{\cal A}(\Omega)$, we get
\begin{equation}
\rm \frac{d^2 {\cal A}}{d\Omega^2} + Q \frac{d{\cal A}}{d\Omega}
+\left( \Lambda_{eff} e^{6\Omega} e^{f(\theta_i) }+ c_1  \right){\cal A} = 0 ,\label{nc00}
\end{equation}
where the constant $\rm c_1$ is the same, as in the commutative
case, and $\rm f(\theta_i)=\mp 3(\theta_4 \eta+\theta_1\nu+\theta_2 \mu)$. The solution for equation (\ref{nc00}) is written with the
generic Bessel function, making the transformation $\rm
\xi=\Lambda_{eff} e^{\mp3(\theta_4 \eta+\theta_1 \nu+\theta_2 \mu )}
e^{6\Omega}$,
\begin{equation}
\rm {\cal A}(\xi)= \rm  \xi^{-\frac{Q}{12}} \, Z_\rho \left(\pm \frac{1}{3}
\xi^{\frac{1}{2}}\right),
\end{equation}
with order $\rm \rho=\frac{1}{6}\sqrt{Q^2 -4 c_1}$.

As in the commutative case, for the simplest factor ordering, $Q=0$,
simplified solutions are obtained, also the same restrictions on $\rm \Lambda_{eff}$ appear.

For case 3, $\rm \Lambda_{eff}\ne0$,  $\gamma=-1$, and factor
ordering $\rm Q\not=0$, the NCWDW equation has the ``noncommutative
potential" $\rm U(\Omega,\Lambda_{eff},\theta)= be^{6[\Omega -
\frac{i\theta_1}{2}\partial_+ -\frac{i\theta_2}{2} \partial_-
-\frac{i\theta_4}{2}\partial_\phi]} $, with $\rm b=384 \pi G
M_{-1}+\Lambda_{eff}$. After doing a separation of variables, using
$\rm \Psi=e^{\pm i\eta \phi}e^{\pm i\nu\beta_+} e^{\pm i\mu\beta_-
}{\cal A}$, and proceeding as before, we make the transformation
$\rm \xi= be^{\mp 3(\theta_4 \eta+\theta_1 \nu+\theta_2 \mu) }
e^{6\Omega}$,  we arrive to the equation
\begin{equation}
\rm \xi^2 \frac{d^2{\cal A}}{d\xi^2} + \left(1+\frac{Q}{6}\right)
\xi\frac{d{\cal A}}{d\xi} +
\frac{1}{36}\left(\xi+\alpha^2\right){\cal A}=0,
\end{equation}
the solution is written with  generic Bessel functions
\begin{equation}
\rm {\cal A}(\xi)=  \xi^{-\frac{Q}{12}} \, Z_{\pm i\rho} \left(\pm
\frac{1}{3} \xi^{\frac{1}{2}}\right),
\end{equation}
with order $\rm \rho=\frac{1}{6}\sqrt{4\alpha^2 -Q^2}$.

\begin{figure}
\includegraphics[width=12cm]{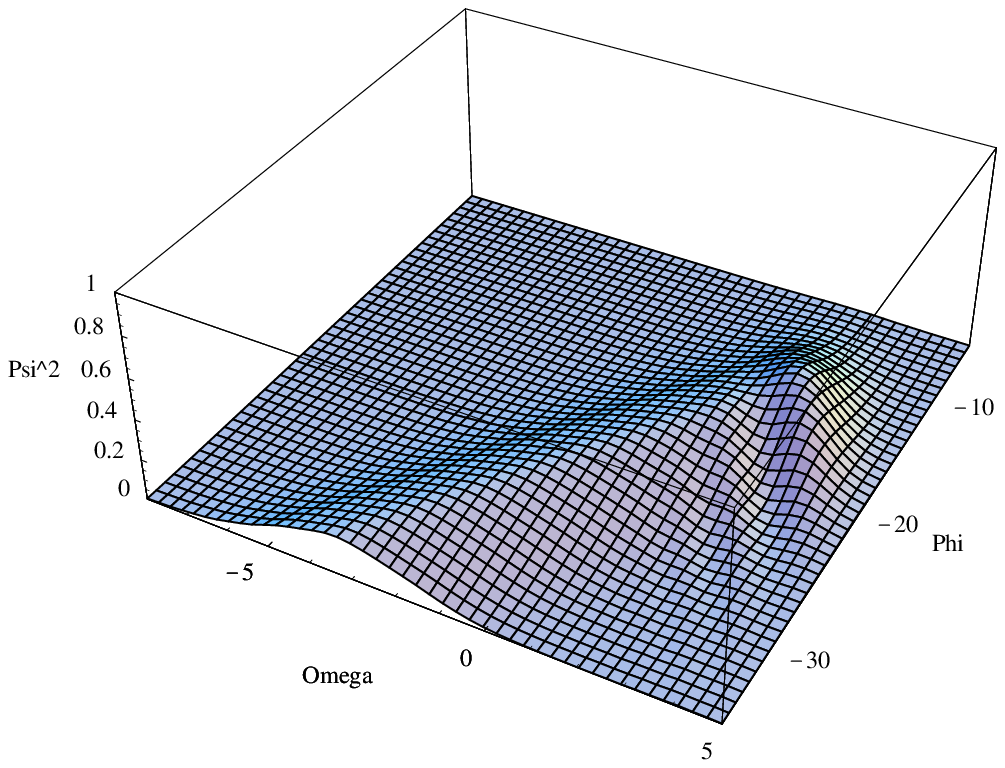}
\caption{\label{phi-gra1} 3D plot for $(\Omega,\phi)$, corresponding to the Gaussian wavepacket equation (\ref{w-p-nc}) 
in the noncommuting case using $\theta=4$and $\mu=\nu=0$. }
\end{figure}

\begin{figure}
\includegraphics[width=12cm]{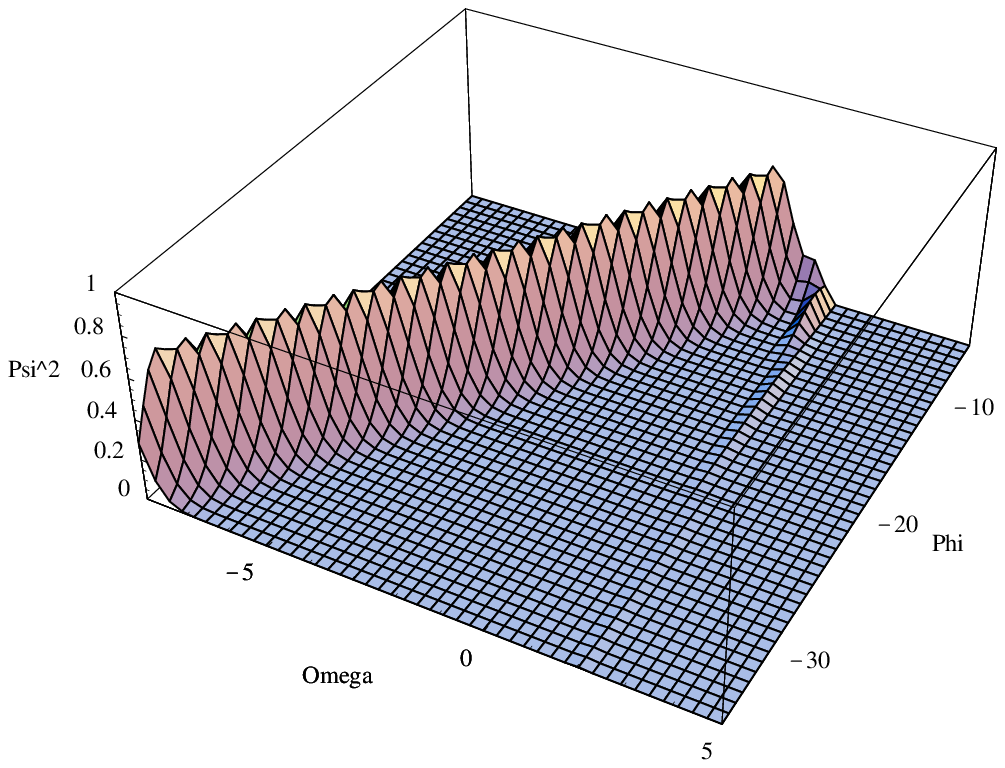}
\caption{\label{phi-gra11} 3D plot for $(\Omega,\phi)$, corresponding to the Gaussian wavepacket equation (\ref{w-p-nc}) 
in the noncommuting case using $\theta=0.5$ and $\mu=\nu=0$. }
\end{figure}

The Gaussian weighted wavepacket  for the particular factor ordering Q=0  is
\begin{equation}
\rm \Psi(\Omega,\phi)={\cal N}\int_{-\infty}^\infty e^{-a(\rho-b)^2}\, e^{-\frac{\sqrt{3}}{2}\rho \phi}\,J_{i\rho}\left(\frac{1}{3}\sqrt{b}
e^{3\Omega} e^{\frac{3\sqrt{3}}{4}\rho \theta_4} \right) d\rho,
\label{w-p-nc}
\end{equation}
 and the  plot for $|\Psi|^2$ can be see in figures (\ref{phi-gra1}-\ref{phi-gra11}) taking two values for 
the noncomutative parameter $\theta=4, 0.5$ respectively. Here again we have dropped the $\beta_\pm$ variables.
 As in the commutative case, we can to change the variable $\phi \to \beta_\pm$, and 
we obtain similar behaviour.

\begin{figure}
\includegraphics[width=12cm]{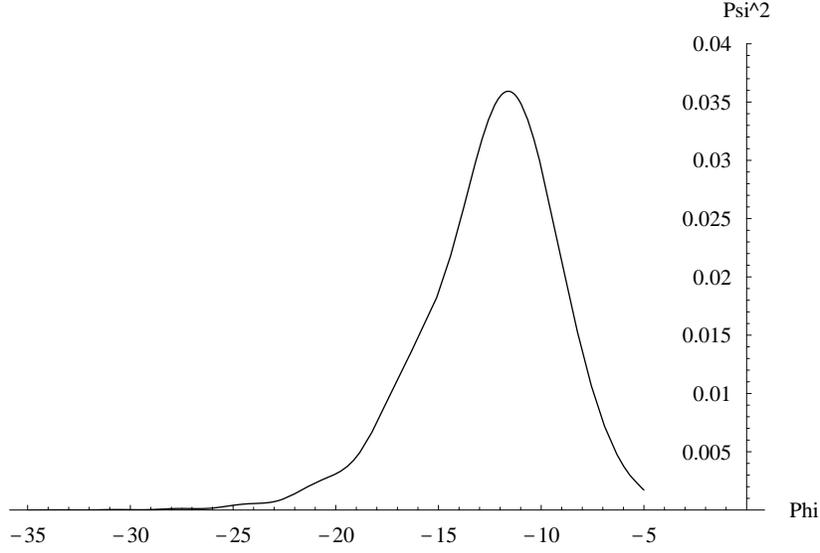}
\caption{\label{phi-gra3} In this plot we present the variation
of $|\Psi|^2$ with respect to $\phi$ when $\theta=0.5$  along the line of $\Omega=1.7$ in figure (\ref{phi-gra11}).}
\end{figure}

\begin{figure}
\includegraphics[width=12cm]{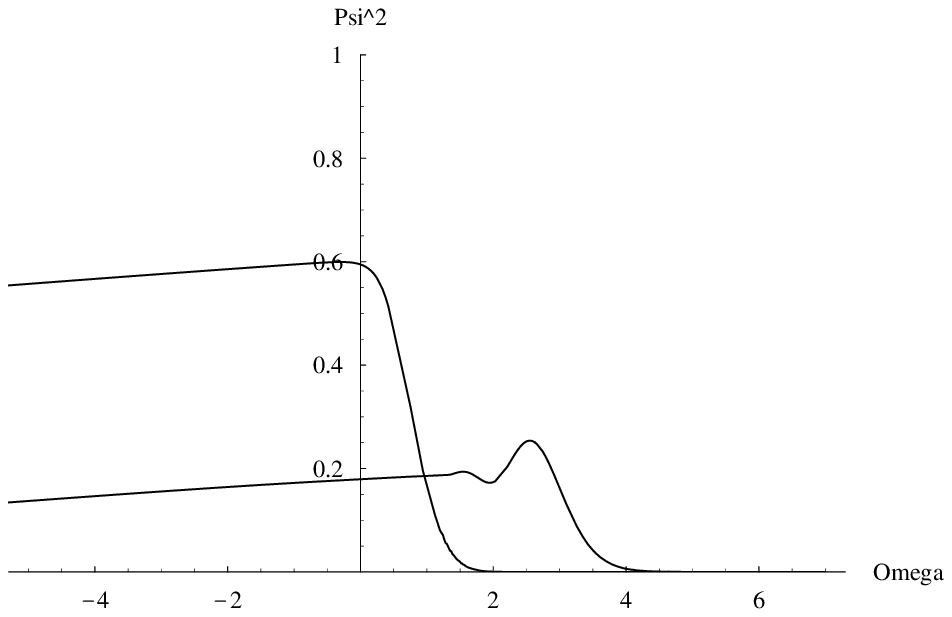}
\caption{\label{phi-gra2} In these plots we present the variation
of $|\Psi|^2$ with respect to $\Omega$ when $\theta=0$ (commutative universe, big plot) and $\theta=4$
(no commutative universe, small plot), along the line of the main structure in figures 1 and 3.}
\end{figure}

 The new structure in the probability density can be see in the figures (\ref{phi-gra3}) and (\ref{phi-gra2}), in the first along the line
of the main structure in figures (1) and (3) when $\theta=4$, second in the line $\Omega=1.7$ presenting a peak in the variable $\phi$.

Finally, for  case $\rm V(\phi)=e^{-\beta \phi}$, stiff fluid $\gamma=1$ and
null cosmological term, the NCWDW becomes
\begin{eqnarray}
   && \rm \left[\frac{\partial^2}{\partial \Omega^2}-
\frac{\partial^2}{\partial \beta_+^2} -
  \frac{\partial^2}{\partial \beta_-^2} -12\frac{\partial^2}{\partial\phi^2}
  +Q\frac{\partial}{\partial \Omega} +384\pi G M_1
  + 12 V_0e^{6[\Omega - \frac{i\theta_1}{2}\partial_+
-\frac{i\theta_2}{2} \partial_--\frac{i\theta_4}{2}\partial_\phi]} \times
\right. \nonumber\\
&&\rm \left. e^{-\beta[\phi+\frac{i\theta_4}{2}\partial_\Omega
-\frac{i\theta_5}{2}\partial_+-\frac{i\theta_6}{2}\partial_-]}
\right]\Psi=0.
\label{wdw2ncq}
\end{eqnarray}
Choosing the wave function $\rm \Psi=e^{\pm i\nu \beta_+} e^{\pm
i\mu \beta_-} {\cal F}(\Omega,\phi)$, we have
\begin{equation}
   \rm \frac{\partial^2 {\cal F}}{\partial \Omega^2} -12\frac{\partial^2 {\cal F}}{\partial\phi^2}  
   +Q\frac{\partial {\cal F}}{\partial \Omega} +\left[\nu^2+ \mu^2+384\pi G M_1
   + 12 V_0 e^{h(\theta_i)}e^{6\Omega-\beta \phi}
e^{-\frac{i\theta_4}{2}[\beta\partial_\Omega+6\partial_\phi]}
\right]{\cal F}=0
\label{wdw2nc}
\end{equation}
where $\rm
h(\theta_i)=(3\theta_1-\frac{\beta}{2}\theta_5)\nu+(3\theta_2-\frac{\beta}{2}\theta_6)\mu$.

It is easier to work in the new variables (x,y) given by (\ref{xy}),
in place of $\rm (\Omega, \phi)$. Thus we have the corresponding
differential equation
\begin{eqnarray}
\rm 12(3-\beta^2)\frac{d^2 {\cal F}}{dx^2} -  6Q \frac{d{\cal F}}{dx} +\left[ 12V_0
e^{h(\theta_i)} e^{-x} + c\right]{\cal F} 
 \rm +\frac{4(\beta^2-3)}{\beta^2}\frac{d^2{\cal F}}{dy^2}
-2Q\frac{d{\cal F}}{dy}=0,\label{xy-nc}
\end{eqnarray}
For solving the equation (\ref{xy-nc}) we choose
the ansatz
$\rm {\cal F}(x,y)=e^{(s-\ell)y} X(x)$, where $\ell=\frac{Q\beta^2}{4(\beta^2-3)}$,
obtaining the following equation for the
function X,
\begin{eqnarray}
 \rm\frac{d^2 X}{dx^2}-\frac{Q}{2(3-\beta^2)}\frac{dX}{dx} + 
+ \rm\frac{1}{12(3-\beta^2)}\left[12V_0 e^{f(\theta_i)} e^{-x}
+n^2\right]X=0, \label{x1}
\end{eqnarray}
where $\rm n^2=c-m^2$, $\rm
m^2=2Q(s-\ell)+\frac{4}{\beta^2}(3-\beta^2)(s-\ell)^2$ and $\rm
f(\theta_i)=h(\theta_i)-i\frac{\theta_4}{\beta}\left(3-\beta^2\right)(s-\ell)$
and s is the constant that appears in the commutative case.

This equation is similar to (\ref{x}), then the solution is similar,
and only suffers a shift in the argument of the Bessel function
\begin{equation}
\rm X(x)=Exp[-\frac{Q}{4(3-\beta^2)}x]J_{\pm r} \left(
\sqrt{\frac{V_0}{3-\beta^2}}e^{f(\theta_i)} e^{-x/2} \right).
\end{equation}
For the remaining case, $\beta^2 = 3$ the solution to the
differential equation is
\begin{equation}
{\cal F}=F_1 e^{-\frac{Q\Omega}{2}}[e^{\frac{6d x +a e^{6x}}{6
\kappa}} e^{-\frac{\kappa y}{4}}].
\label{beta}
\end{equation}
here $\kappa$ is a non null separation constant, $F_1$ is an
arbitrary integration constant, $a=c-Q^2/4$, $d= 12V_0e^{m(\theta_i)}=b e^{m(\theta_i)}$,with
\begin{equation}
m(\theta_i)=(3\theta_1\mp
\frac{\sqrt{3}}{2}\theta_5)\nu+(3\theta_2\mp
\frac{\sqrt{3}}{2}\theta_6)\mu\mp \frac{\sqrt{3}}{4}\theta_4 \kappa.
\end{equation}
and the  variables $x$ and $y$ are defined as in the commutative
case (equation (\ref{tranf})) .

\section{Final remarks}
In this work by using the equivalence  between General Relativity 
 and Hamiltonian formalism,  noncommutative scenarios are constructed.
 This was achieved by deforming the minisuperspace  for the Bianchi type I cosmological 
 model coupled to  barotropic perfect fluid, field $\phi$ and cosmological term. 
 in the gauge $\rm N=1$, as we can see
 the solution $\rm \Omega_{nc}$ is  the commutative solution  plus a function on
 $\theta_i$, independent of time.
 We have also analyzed the $\rm \beta_{\pm nc}$ 
 noncommutative solutions, in some ranges on the parameter
 $\theta_i$ and effective cosmological constant, occurs a  dynamical isotropization,
 i.e.,  $\beta_{\pm nc}$ are bounded in the present time, noting  that different evolution 
 with respect to the commutative $\beta_\pm$, also 
 the field $\phi$ decays dynamically to  a constant, 
 but after a critical time $t_c$ the evolution is the same in both scenarios. 
In the quantum stadium, in all but one of the cases considered the influence of the noncommutativity is 
encoded  in a change of the argument of Bessel functions and the change is not 
qualitatively very remarkable ( see figures (\ref{phi-gra},\ref{phi-gra1},\ref{phi-gra2}). 
The singular case is the one with $\beta^2=3$ and 
here the influence of the noncommutativity is in the argument of an exponential in 
the form $ 6 d x + a e^{6x}$. 
Only the  constant d depends on the noncommutivity parameters and therefore we 
have that for $x$
large and negative they will dominate and for large positive $x$ the noncommutative 
effects disappear.
 From the definition of $x$  we see that for finite $\phi$ the noncommutative regime corresponds 
 to $\Omega \to -\infty $ and at late times commutativity is recovered.   It would be of interest to improve the work with the two
 approaches cited in the references \cite{gup,dsr},  and obtain the possibles differences in the dynamics of our universe, using anisotropic
 cosmological models; this will be reported in future work.

\acknowledgments{This work was supported in part by CONACyT grant 47641, DINPO, CONCYTEG and
Promep grant UGTO-CA-3. Many calculations where done by Symbolic Program REDUCE 3.8. We also like to
thank M. Sabido for usefull comments. This work is part of the collaboration within the Instituto Avanzado de
Cosmolog\'{\i}a.}

\end{document}